\documentclass[%
]{algotel}
\usepackage[latin1]{inputenc}
\usepackage[francais]{babel}
\usepackage{tikz}
\usepackage{pgfplots}
\usepackage{subfigure} 

\author{Yoann Pigné\addressmark{1}, Arnaud Casteigts\addressmark{2}, Frédéric Guinand\addressmark{3}, et Serge Chaumette\addressmark{4}}

\title[Construction et maintien d'une forêt couvrante dans un réseau dynamique]{Construction et maintien d'une forêt couvrante dans un réseau dynamique}

\address{\addressmark{1} SNT - Université du Luxembourg \\
         \addressmark{2} SITE - Université d'Ottawa, Canada \\
         \addressmark{3} LITIS - Université du Havre, France\\
         \addressmark{4} LaBRI - Université Bordeaux 1, France \\}


\begin{document}
\maketitle

\begin{abstract}
Dans ce travail nous présentons le principe d'un algorithme de construction et de maintien d'une forêt couvrante dans un réseau de télécommunication mobile de type réseau mobile ad hoc ({\it MANET}). L'algorithme, basé sur une marche aléatoire de jetons, est entièrement décentralisé. Nous en proposons une analyse probabiliste dans le cadre statique et nous montrons comment l'ajout d'une mémoire aux jetons permet d'en améliorer sensiblement les performances.
\end{abstract}

\section{Introduction}
\label{sec:in}

Les arbres couvrants peuvent rendre de nombreux services dans les réseaux de télécommunications, en simplifiant notamment les tâches les plus courantes telles que la diffusion d'information, le routage ou l'exclusion mutuelle. 
Dans les réseaux statiques ou faiblement dynamiques, une distinction est généralement faite entre l'étape de construction d'un tel arbre, et celle de son utilisation effective (la seconde succédant à la première de manière disjointe). Malheureusement, cette approche n'est pas adaptée aux réseaux fortement dynamiques dans lesquels aucune période de stabilité ne peut être supposée, et où les changements affectant la topologie du réseau sont susceptibles de partionner ce dernier en plusieurs composantes connexes. 
Dans un tel contexte, la gestion de la structure couvrante doit être vue comme un processus continu et 
perpétuel, alternant ruptures d'arbres existants et fusions de nouveaux arbres (d'où les termes de {\em maintien}, et de {\em forêt} d'arbres couvrants) \cite{Baala2003}. 

Nous nous intéressons ici à étudier ce qu'il est possible de construire dans ce cadre particulier,
en éliminant également l'hypothèse selon laquelle les stations sont dotées d'identifiants uniques.
L'approche que nous proposons repose sur une marche aléatoire de jetons. L'idée sous-jacente est relativement simple~: chaque arbre dans la forêt possède exactement un jeton dont la marche est strictement limitée à ses arêtes. Lorsque deux jetons se rencontrent (sont positionnés sur deux sommets voisins dans le graphe des connexions), les deux arbres associés aux jetons fusionnent de manière locale et atomique, et l'un des jetons est supprimé. Lorsqu'une arête d'un arbre disparaît, l'arbre est scindé en deux sous-arbres et le sommet adjacent à cette arête appartenant au sous-arbre sans jeton en génère un nouveau (là encore, de manière locale et atomique). 

Après avoir détaillé le fonctionnement de l'algorithme (originellement présenté dans~\cite{casteigts_06}) dans la Section~\ref{sec:algo}, nous proposons une version optimisée pour laquelle la marche des jetons n'est plus tout à fait aléatoire, mais devient en partie contrainte par ses mouvements passés.
Nous présentons ensuite quelques résultats préliminaires d'analyse probabiliste ainsi que nos simulation éclairant les performances absolues et relatives de ces deux variantes dans un contexte statique. L'élaboration d'un cadre d'étude approprié à leur analyse en contexte dynamique fait l'objet de travaux actuels, et est laissé ici ouvert.

\bigskip

\section{Algorithmes}
\label{sec:algo}

En préambule à cette partie, considérons un graphe $G=(S,A)$ couvert par deux arbres $T_1=(S_1,A_1)$ et $T_2=(S_2,A_2)$. Soient $u$ et $v$ deux sommets tels que $u \in S_1$ et$v \in S_2$ tels que $(u,v) \in A$. D'un point de vue décentralisé, le problème qui se pose à ces sommets et de savoir si leur arête commune doit être utilisée pour fusionner leur arbre respectif. Pour prendre la bonne décision, deux problèmes doivent être résolus~: (1) est-ce que $u$ et $v$ appartiennent à deux arbres différents~? Ou bien, existe-t-il un chemin dans l'arbre liant ces deux sommets~? (2) Comment garantir qu'aucune autre opération de fusion des deux arbres est en cours entre deux autres sommets situés ailleurs dans l'arbre~?

Le second problème implique que i) avant de fusionner un sommet sait qu'il est le seul et unique capable d'effectuer une telle opération dans son arbre à cette date, ou que ii)  le sommet initie une consultation dans son arbre pour obtenir des autres sommets ou d'une autorité centrale, la permission d'effectuer l'opération de fusion. De manière pratique, dans un réseau dynamique, le lancement d'une consultation ne peut pas être envisagé à moins de considérer des hypothèses fortes sur les délais séparant deux séries de changements de la topologie. La meilleure option semble être la conception d'un mécanisme qui permet une prise de décision locale en autorisant un  seul et unique sommet à déclencher une opération de fusion. 

\subsection{L'algorithme de référence}

Les règles définissant l'algorithme sont présentées dans la Figure~\ref{fig:algo1}.

\vspace{-0.3cm}

\begin{center}
\begin{figure}[h]
\noindent
{\footnotesize \underline{état initial~:} \texttt{T} pour chaque sommet, {\scriptsize $\emptyset$} pour chaque extrémité d'arête.}\\
\begin{tikzpicture}[scale=1]
  \tikzstyle{every node}=[draw,circle,fill=black!80,inner sep=1.8pt]
  \path (-1.5,0) node (v11) {};
  \path (1.3,0) node (v21) {};
  \tikzstyle{every node}=[]
  \path (-.7,0) node (v12) {};
  \draw[thick, ->, shorten >=10pt, shorten <=10pt] (v12)--(v21);
  \draw (v11)--(v12);
  \tikzstyle{every node}=[font=\footnotesize]
  \path (-2,0) node (rb) {$r_1:$ };
  \path (v11.north)+(0,.15) node (lab) {\texttt{N}};
  \path (v21.north)+(0,.15) node (lab) {\texttt{T}};
  \tikzstyle{every node}=[font=\tiny]
  \path (v11.east)+(.12,-.12) node (lab) {\texttt{1}};
  \path (v11.east)+(.4,.12) node (lab) {off};
 \tikzstyle{every node}=[draw,circle,fill=black!80,inner sep=1.8pt]
  \path (5.5,0) node (u11) {};
  \path (7.3,0) node (u21) {};
  \tikzstyle{every node}=[]
  \path (5.3,0) node (u12) {};
  \draw[thick, ->, shorten >=10pt, shorten <=10pt] (u12)--(u21);
  \draw (u11)--(u12);
  \tikzstyle{every node}=[font=\footnotesize]
  \path (5,0) node (rb) {$r_2:$ };
  \path (u11.north)+(0,.15) node (lab) {Any};
  \path (u21.north)+(0,.15) node (lab) {Any};
  \tikzstyle{every node}=[font=\tiny]
  \path (u11.east)+(0.12,-.12) node (lab) {\texttt{2}};
  \path (u11.east)+(0.4,.12) node (lab) {off};
\end{tikzpicture}\\
\begin{tikzpicture}[scale=1]
  \tikzstyle{every node}=[draw,circle,fill=black!80,inner sep=1.8pt]
  \path (-1.5,0) node (v11) {};
  \path (-.2,0) node (v12) {};
  \path (1.7,0) node (v21) {};
  \path (3,0) node (v22) {};
  \draw[thick, ->, shorten >=10pt, shorten <=10pt] (v12)--(v21);
  \draw (v11)--(v12);
  \draw (v21)--(v22);
  \tikzstyle{every node}=[font=\footnotesize]
  \path (-2,0) node (rb) {$r_3:$ };
  \path (v11.north)+(0,.15) node (lab) {\texttt{T}};
  \path (v12.north)+(0,.15) node (lab) {\texttt{T}};
  \path (v21.north)+(0,.15) node (lab) {\texttt{T}};
  \path (v22.north)+(0,.15) node (lab) {\texttt{N}};
  \tikzstyle{every node}=[font=\tiny]
  \path (v11.east)+(.12,-.12) node (lab) {$\emptyset$};
  \path (v12.west)+(-.12,-.12) node (lab) {$\emptyset$};
  \path (v21.east)+(.12,-.12) node (lab) {\texttt{2}};
  \path (v22.west)+(-.12,-.12) node (lab) {\texttt{1}};
  \tikzstyle{every node}=[draw,circle,fill=black!80,inner sep=1.8pt]
  \path (5.5,0) node (v11) {};
  \path (6.8,0) node (v12) {};
  \path (8.7,0) node (v21) {};
  \path (10,0) node (v22) {};
  \draw[thick, ->, shorten >=10pt, shorten <=10pt] (v12)--(v21);
  \draw (v11)--(v12);
  \draw (v21)--(v22);
  \tikzstyle{every node}=[font=\footnotesize]
  \path (5,0) node (rb) {$r_4:$ };
  \path (v11.north)+(0,.15) node (lab) {\texttt{T}};
  \path (v12.north)+(0,.15) node (lab) {\texttt{N}};
  \path (v21.north)+(0,.15) node (lab) {\texttt{N}};
  \path (v22.north)+(0,.15) node (lab) {\texttt{T}};
  \tikzstyle{every node}=[font=\tiny]
  \path (v11.east)+(.12,-.12) node (lab) {\texttt{2}};
  \path (v12.west)+(-.12,-.12) node (lab) {\texttt{1}};
  \path (v21.east)+(.12,-.12) node (lab) {\texttt{1}};
  \path (v22.west)+(-.12,-.12) node (lab) {\texttt{2}};
\end{tikzpicture}
\caption{L'algorithme de construction et de maintien d'une forêt couvrante sous la forme d'un ensemble de règle de réétiquetage.}
\label{fig:algo1}
\end{figure}
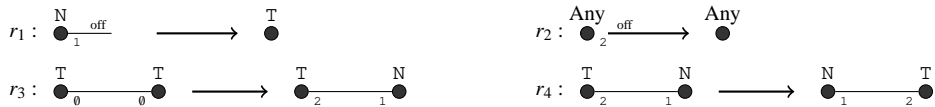
\end{center}

\vspace{-0.9cm}

L'algorithme est basé sur le mouvement des jetons dans leur arbre respectif. Les jetons sont autorisés à effectuer trois opérations~: la \textit{circulation}, la \textit{fusion}, et la \textit{régénération}, dont l'objectif est de maintenir un, et seulement un, jeton par arbre. Initialement, chaque sommet est un arbre en soit qui possède son propre jeton (étiqueté \texttt{T}). Lorsque deux jetons sont voisins, c'est-à-dire sont situés sur des sommets à distance 1 l'un de l'autre, ils fusionnent en un seul jeton et les deux arbres fusionnent pour n'en former qu'un seul. Le sommet qui conserve le jeton reste étiqueté \texttt{T} tandis que le second est réétiqueté \texttt{N}, et l'arête est marquée comme une {\em arête d'arbre} (règle~$r_3$) en utilisant une étiquette différente sur chaque extrémité (\texttt{1} et \texttt{2}) pour rendre compte de l'orientation induite par la position du jeton restant.  Lorsqu'aucune fusion n'est possible, le jeton est transmis à un sommet voisin \underline{dans} l'arbre (règle ~$r_4$), et le marquage de l'orientation est mis à jour en conséquence.
Le point central est que les étiquettes des arêtes définissent toujours un arbre orienté dont la racine est le sommet qui contient le jeton. Grâce à ces étiquettes, chaque sommet étiqueté \texttt{N} possède une unique arête sortante (étiquette $1$), ce qui donne la direction vers le jeton. Si une telle arête disparaît pour une quelconque raison, alors ce sommet sait qu'il est maintenant en position de racine pour le sous-arbre qui ne possède plus de jeton. Il est alors en mesure de régénérer un nouveau jeton (règle $r_1$). Cette propriété reste vraie quelque soit le nombre de disparitions simultanées d'arêtes dans l'arbre. Pour le sommet situé à l'autre extrémité de l'arête, la disparition de l'arête induit simplement la suppression de l'état local de l'arête perdue (règle $r_2$), indépendamment du fait que ce sommet ait ou non le jeton. Pour favoriser les fusions potentielles, nous considérons que l'application des règles suit un ordre, c'est-à-dire que la règle $r_4$ ne peut pas être appliquée si la règle $r_3$ est applicable, ce qui suppose qu'un sommet étudie l'ensemble de son voisinage avant l'application d'une règle.

\subsection{Analyse des performances}
\label{sec:analysealgoreference}


Considérons un jeton situé sur un sommet $u$ qui ne soit pas en mesure d'appliquer la règle $r_3$ (fusion). Le sommet applique alors la règle $r_4$ de transmission du jeton vers l'un de ses voisins. Le choix de ce voisin est effectué de manière équiprobable, ainsi le déplacement du jeton dans l'arbre $\mathcal{T}$ est une {\em marche aléatoire} dans cet arbre. Partant de ce fait, la probabilité pour le jeton d'être situé sur un sommet $v$ dépend uniquement du degré $d_{\mathcal{T}}(v)$ de ce sommet et du nombre $n$ de sommets de l'arbre~:
 
\begin{equation}
  \label{eq:proba-each}
 P(\mbox{jeton présent sur $v$}) = \frac{d_{\mathcal{T}}(v)}{2(n-1)} 
\end{equation}

Pour évaluer la performance de cet algorithme, nous considérons la {\em distribution stationnaire} vers laquelle tend la marche aléatoire des jetons. Il nous faut cependant considérer deux hypothèses. La première concerne la convergence de la marche aléatoire vers la distribution stationnaire qui n'est vraie que pour des graphes non bipartis. Or nos graphes sont des arbres. Nous évacuons ce problème en considérant que les mouvements des jetons dans les différents arbres ne sont pas synchronisés. La seconde hypothèse tient au temps nécessaire à la marche aléatoire pour converger vers la distribution stationnaire, ce temps, {\em le mixing time}, n'est pas toujours négligeable, pourtant, nous le considérerons comme tel par la suite. 

Nous sommes intéressés par le temps de fusion, c'est-à-dire le nombre de mouvements nécessaires aux jetons pour pouvoir fusionner. Considérons un graphe $G=(S,A)$ et deux arbres $\mathcal{T}_1=(V_{\mathcal{T}_1},E_{\mathcal{T}_1})$ et $\mathcal{T}_2=(V_{\mathcal{T}_2},E_{\mathcal{T}_2})$ sous-graphes du graphe $G$. Soit une arête $(u,v) \in A$ telle que $u \in \mathcal{T}_1$ et $v \in \mathcal{T}_2$, une telle arête est qualifiée de {\em pont}. 

Le {\em temps de fusion estimé} est une estimation du nombre moyen d'étapes (nombre de mouvements des jetons dans le contexte qui nous occupe) requis pour fusionner deux arbres. Ce temps peut être calculé à partir de la probabilité que les deux jetons soient à distance 1 l'un de l'autre dans le graphe. Or, cette probabilité se calcule simplement à partir de l'équation \ref{eq:proba-each}~:

\begin{equation}
  \label{eq:proba-mixed}
  \footnotesize 
  P_{\mbox{fusion}}(\mathcal{T}_1,\mathcal{T}_2)=\sum_{\{(u,v) \in \mbox{Ponts}(\mathcal{T}_1,\mathcal{T}_2)\}} \frac{d_{\mathcal{T}_1}(u)}{2|E_{\mathcal{T}_1}|} \times \frac{d_{\mathcal{T}_2}(v)}{2|E_{\mathcal{T}_2}|}
\end{equation} 

Ainsi, le nombre moyen de déplacements de jetons nécessaires pour la fusion de deux arbres $\mathcal{T}_1$ et $\mathcal{T}_2$ vaut~: 

\begin{equation}
  \label{eq:proba-expect}
  \footnotesize 
  E_{\mbox{fusion}}(\mathcal{T}_1,\mathcal{T}_2)= \left({P_{\mbox{fusion}}(\mathcal{T}_1,\mathcal{T}_2)}\right)^{-1} =  \left(\sum_{\{(u,v) \in \mbox{Ponts}(\mathcal{T}_1,\mathcal{T}_2)\}} \frac{d_{\mathcal{T}_1}(u)}{2|E_{\mathcal{T}_1}|} \times \frac{d_{\mathcal{T}_2}(v)}{2|E_{\mathcal{T}_2}|}\right)^{-1}
\end{equation}

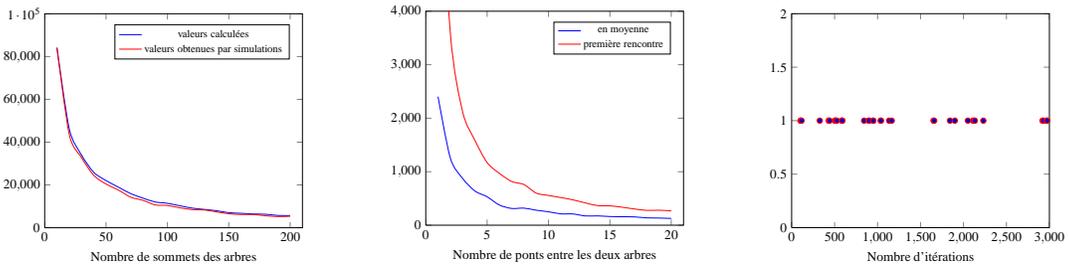
\begin{figure}[!ht]
  \begin{center}
    \subfigure[\small\emph{Comparaison entre nombre de rencontres calculées selon la formule \ref{eq:proba-expect} (courbe bleue) et valeurs obtenues par simulation (courbe rouge)}]
    {\label{fig:comparaisonanalytiquerw} 
      \begin{tikzpicture}[scale=.5]
        \pgfplotsset{every axis y label/.style={at={(-0.06,0.5)}, xshift=-10pt, rotate=90}}
        \pgfplotsset{every axis legend/.append style={at={(0.95,.95)},anchor=north east}}
        \begin{axis}[xlabel={Nombre de sommets des arbres},ylabel={},xmode=normal,ymin=0,ymax=100000,xmin=0,xmax=210,scaled ticks=false]
          \addlegendentry{\footnotesize valeurs calculées}
          \addplot[smooth,blue] plot coordinates {
            (10,84130) 
            (20,46279) 
            (30,34024) 
            (40,25884) 
            (50,22053) 
            (60,18990) 
            (70,15964) 
            (80,13861) 
            (90,12044) 
            (100,11416)
            (110,10283)
            (120,9175) 
            (130,8568) 
            (140,7946) 
            (150,7047) 
            (160,6750) 
            (170,6435) 
            (180,6304) 
            (190,5739) 
            (200,5702) 
          };
          \addlegendentry{\footnotesize valeurs obtenues par simulations}
          \addplot[smooth,red] plot coordinates {
            (10,84158)
            (20,43651)
            (30,32807)
            (40,24469)
            (50,20421)
            (60,17645)
            (70,14299)
            (80,12812)
            (90,10669)
            (100,10400)
            (110,9320)
            (120,8480)
            (130,8312)
            (140,7412)
            (150,6531)
            (160,6182)
            (170,6163)
            (180,5584)
            (190,5226)
            (200,5349)
          };
        \end{axis}
      \end{tikzpicture}~~~~~
    }
    \hspace{0.3cm}
    \subfigure[\small\emph{Mesure de la différence entre nombre de mouvements moyen pour une rencontre et nombre moyen pour la première rencontre (simulation)}]
     {\label{fig:compmeanfirst}

       \begin{tikzpicture}[scale=.5]
         \pgfplotsset{every axis y label/.style={at={(-0.06,0.5)}, xshift=-10pt, rotate=90}}
         \pgfplotsset{every axis legend/.append style={at={(0.95,.95)},anchor=north east}}
         \begin{axis}[xlabel={Nombre de ponts entre les deux arbres},ylabel={},xmode=normal,ymin=0,ymax=4000,xmin=0,xmax=21,scaled ticks=false]
           \addlegendentry{\footnotesize en moyenne}
           \addplot[smooth,blue] plot coordinates {
             (1, 2403)
             (2, 1256)
             (3, 875)
             (4, 640)
             (5, 531)
             (6, 377)
             (7, 313)
             (8, 319)
             (9, 280)
             (10, 251)
             (11, 212)
             (12, 210)
             (13, 173)
             (14, 174)
             (15, 162)
             (16, 158)
             (17, 155)
             (18, 137)
             (19, 134)
             (20, 126)
           };
           \addlegendentry{\footnotesize première rencontre}
           \addplot[smooth,red] plot coordinates {
             (1, 7637)
             (2,3595)
             (3,2124)
             (4,1589)
             (5,1170)
             (6,968)
             (7,817)
             (8,760)
             (9,600)
             (10,556)
             (11,517)
             (12,473)
             (13,417)
             (14,366)
             (15,362)
             (16,336)
             (17,303)
             (18,277)
             (19,280)
             (20,267)
           };
         \end{axis}
       \end{tikzpicture}~~~~~
     }
    \hspace{0.3cm}
    \subfigure[\small\emph{Distribution des passages du jeton sur un sommet (choisi aléatoirement) de l'arbre formé de 50 sommets (simulation)}]
     {\label{fig:rwpassage}
       \begin{tikzpicture}[scale=.5]
         \pgfplotsset{every axis y label/.style={at={(-0.06,0.5)}, xshift=-10pt, rotate=90}}
         \pgfplotsset{every axis legend/.append style={at={(0.95,.95)},anchor=north east}}
         \begin{axis}[xlabel={Nombre d'itérations},ylabel={},xmode=normal,ymin=0,ymax=2,xmin=0,xmax=3000,scaled ticks=false]
           \addplot+[only marks,red] plot coordinates {
             (103,1)
             (105,1)
             (115,1)
             (327,1)
             (431,1)
             (437,1)
             (441,1)
             (501,1)
             (509,1)
             (515,1)
             (527,1)
             (585,1)
             (587,1)
             (589,1)
             (843,1)
             (895,1)
             (903,1)
             (949,1)
             (951,1)
             (1037,1)
             (1039,1)
             (1135,1)
             (1165,1)
             (1651,1)
             (1659,1)
             (1843,1)
             (1899,1)
             (2051,1)
             (2111,1)
             (2127,1)
             (2131,1)
             (2231,1)
             (2919,1)
             (2931,1)
             (2969,1)
             (2971,1)
             (2973,1)
           };
         \end{axis}
       \end{tikzpicture}~~~~~
     }
     \caption{\small\emph{Résultats obtenus pour l'étude de la rencontre de deux jetons animés d'une marche aléatoire non biaisée.}\label{fig:rw}}
  \end{center}
\end{figure}

On constate sur la figure \ref{fig:comparaisonanalytiquerw} que les valeurs obtenues par le calcul analytique et les valeurs obtenues par simulation correspondent parfaitement pour des arbres dont l'ordre varie entre 10 et 200. 

\vspace{-0.3cm}

Cependant, il s'agit de valeurs moyennes, or dans le contexte de cette étude, le nombre moyen de mouvements nécessaires pour que les deux jetons soient en position d'appliquer la règle $r_3$ n'est pas pertinent, ce qui nous intéresse est le nombre de mouvements nécessaires pour qu'ait lieu la {\em première} rencontre entre les deux jetons. Or on constate une importante dérive entre cette dernière valeur et la valeur moyenne comme l'illustre la figure \ref{fig:compmeanfirst}. 

Si l'on effectue une analyse du déplacement d'un jeton dans un arbre, et que l'on s'intéresse aux passages de ce jeton sur un sommet particulier, la distribution temporelle des passages du jeton sur le sommet ressemble à la distribution illustrée par la Figure \ref{fig:rwpassage}. En partant de l'hypothèse que cette distribution est représentative, nous pouvons en déduire qu'en moyenne, lorsque le jeton est situé à une distance importante du sommet cible, un nombre important de mouvements est nécessaire pour que le jeton revienne dans le voisinage du sommet cible.
Dans le contexte de la rencontre de deux jetons se déplaçant au sein de deux arbres, il semble que ce comportement ait un impact négatif sur les performances de l'algorithme. Une solution consisterait à améliorer la fluidité du mouvement du jeton dans l'arbre. C'est l'objet de la section suivante.

\subsection{Optimisation de l'algorithme}
\label{sec:algoopt}

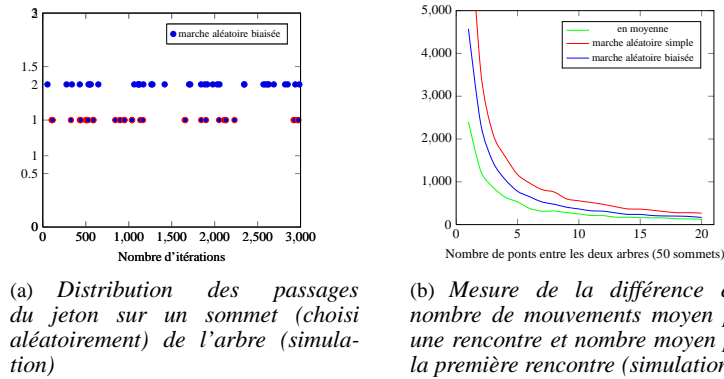
\begin{figure}[!ht]
  \begin{center}
    \subfigure[\small\emph{Distribution des passages du jeton sur un sommet (choisi aléatoirement) de l'arbre (simulation)}]
     {\label{fig:rwpassage2}
       \begin{tikzpicture}[scale=.5]
         \pgfplotsset{every axis y label/.style={at={(-0.06,0.5)}, xshift=-10pt, rotate=90}}
         \pgfplotsset{every axis legend/.append style={at={(0.95,.95)},anchor=north east}}
         \begin{axis}[xlabel={Nombre d'itérations},ylabel={},xmode=normal,ymin=0,ymax=2,xmin=0,xmax=3000,scaled ticks=false]
           \addlegendentry{\footnotesize marche aléatoire simple}
           \addplot+[only marks,red] plot coordinates {
             (103,1)
             (105,1)
             (115,1)
             (327,1)
             (431,1)
             (437,1)
             (441,1)
             (501,1)
             (509,1)
             (515,1)
             (527,1)
             (585,1)
             (587,1)
             (589,1)
             (843,1)
             (895,1)
             (903,1)
             (949,1)
             (951,1)
             (1037,1)
             (1039,1)
             (1135,1)
             (1165,1)
             (1651,1)
             (1659,1)
             (1843,1)
             (1899,1)
             (2051,1)
             (2111,1)
             (2127,1)
             (2131,1)
             (2231,1)
             (2919,1)
             (2931,1)
             (2969,1)
             (2971,1)
             (2973,1)
           };
         \end{axis}
         \begin{axis}[xlabel={Nombre d'itérations},ylabel={},xmode=normal,ymin=0,ymax=3,xmin=0,xmax=3000,scaled ticks=false]
           \addlegendentry{\footnotesize marche aléatoire biaisée}
           \addplot+[only marks,blue] plot coordinates {
             (55,2)
             (277,2)
             (339,2)
             (431,2)
             (533,2)
             (547,2)
             (559,2)
             (647,2)
             (1065,2)
             (1107,2)
             (1121,2)
             (1125,2)
             (1171,2)
             (1265,2)
             (1277,2)
             (1419,2)
             (1703,2)
             (1717,2)
             (1843,2)
             (1885,2)
             (1915,2)
             (1979,2)
             (2031,2)
             (2057,2)
             (2343,2)
             (2347,2)
             (2569,2)
             (2591,2)
             (2617,2)
             (2629,2)
             (2689,2)
             (2821,2)
             (2851,2)
             (2939,2)
             (2987,2)
           };
         \end{axis}
       \end{tikzpicture}~~~~~
     }
     \hspace{0.5cm}
     \subfigure[\small\emph{Mesure de la différence entre nombre de mouvements moyen pour une rencontre et nombre moyen pour la première rencontre (simulation)}]
       {\label{fig:compmeanfirsttabou}
       \begin{tikzpicture}[scale=.5]
         \pgfplotsset{every axis y label/.style={at={(-0.06,0.5)}, xshift=-10pt, rotate=90}}
         \pgfplotsset{every axis legend/.append style={at={(0.95,.95)},anchor=north east}}
         \begin{axis}[xlabel={Nombre de ponts entre les deux arbres (50 sommets)},ylabel={},xmode=normal,ymin=0,ymax=5000,xmin=0,xmax=21,scaled ticks=false]
           \addlegendentry{\footnotesize en moyenne}
           \addplot[smooth,green] plot coordinates {
             (1, 2403)
             (2, 1256)
             (3, 875)
             (4, 640)
             (5, 531)
             (6, 377)
             (7, 313)
             (8, 319)
             (9, 280)
             (10, 251)
             (11, 212)
             (12, 210)
             (13, 173)
             (14, 174)
             (15, 162)
             (16, 158)
             (17, 155)
             (18, 137)
             (19, 134)
             (20, 126)
           };
           \addlegendentry{\footnotesize marche aléatoire simple}
           \addplot[smooth,red] plot coordinates {
             (1, 7637)
             (2,3595)
             (3,2124)
             (4,1589)
             (5,1170)
             (6,968)
             (7,817)
             (8,760)
             (9,600)
             (10,556)
             (11,517)
             (12,473)
             (13,417)
             (14,366)
             (15,362)
             (16,336)
             (17,303)
             (18,277)
             (19,280)
             (20,267)
           };
           \addlegendentry{\footnotesize marche aléatoire biaisée}
           \addplot[smooth,blue] plot coordinates {
             (1,4575)
             (2,2358)
             (3,1464)
             (4,1051)
             (5,779)
             (6,650)
             (7,528)
             (8,475)
             (9,406)
             (10,366)
             (11,322)
             (12,314)
             (13,274)
             (14,238)
             (15,239)
             (16,211)
             (17,202)
             (18,199)
             (19,190)
             (20,167)
           };
         \end{axis}
       \end{tikzpicture}~~~~~
     }
     \caption{\small\emph{Résultats obtenus pour l'étude de la rencontre de deux jetons animés d'une marche aléatoire biaisée par un mécanisme restreignant ses choix.}\label{fig:tabou}}
  \end{center}
\end{figure}

Pour améliorer les performances de l'algorithme, c'est-à-dire pour réduire les temps de fusion des arbres, nous pensons qu'il peut être pertinent de fluidifier le mouvement du jeton dans l'arbre de manière à ce que la distribution de ses passages sur les sommets se rapproche d'une distribution uniforme. Pour cela, nous proposons de restreindre les choix du jeton pour son prochain déplacement, afin que soient privilégiées les parties de l'arbre qui n'ont pas été visitées récemment. Le principe repose sur l'interdiction faite au jeton d'effectuer le mouvement inverse du dernier mouvement accompli, sauf si le sommet sur lequel le jeton est positionné est une feuille. Les résultats obtenus sont présentés dans la Figure \ref{fig:tabou}. Il apparaît clairement que cette simple modification de l'algorithme permet d'améliorer notablement les temps de fusion des arbres. Des expériences sont actuellement en cours pour mesurer l'impact de cette amélioration dans le contexte dynamique. L'ensemble des simulations réalisées pour ce travail ont été effectuées en utilisant la librairie GraphStream \cite{epnacs2007}.

\vspace{-0.6cm}

\nocite{*}
\bibliographystyle{alpha}
\bibliography{algotel}
\label{sec:biblio}

\end{document}